\documentclass[twocolumn,aps,prl,amsmath,amssymb]{revtex4-1}
\usepackage[latin9]{inputenc}
\usepackage{mathrsfs}
\usepackage{amsmath}
\usepackage{amssymb}
\usepackage{graphicx}
\usepackage{esint}

\makeatletter

\newcommand*\LyXThinSpace{\,\hspace{0pt}}

\usepackage{epsfig}
\usepackage{bm}
\usepackage{dcolumn}
\usepackage{color}

\makeatother

\begin{document}
\title{Consistent theory of self-bound quantum droplets with bosonic pairing }
\author{Hui Hu and Xia-Ji Liu}
\affiliation{Centre for Quantum Technology Theory, Swinburne University of Technology,
Melbourne, Victoria 3122, Australia}
\date{\today}
\begin{abstract}
We revisit the Bogoliubov theory of quantum droplets proposed by Petrov
{[}Phys. Rev. Lett. \textbf{115}, 155302 (2015){]} for an ultracold
Bose-Bose mixture, where the mean-field collapse is stabilized by
the Lee-Huang-Yang quantum fluctuations. We show that a loophole in
Petrov's theory, i.e., the ignorance of the softening complex Bogoliubov
spectrum, can be naturally removed by the introduction of bosonic
pairing. The pairing leads to weaker mean-field attractions, and also
stronger Lee-Huang-Yang term in the case of unequal intraspecies interactions.
As a result, the equilibrium density for the formation of self-bound
droplets significantly decrease in the deep droplet regime, in agreement
with a recent observation from diffusion Monte Carlo simulations.
Our construction of a consistent Bogoliubov theory paves the way to
understand the puzzling low critical number of small quantum droplets
observed in the experiment {[}Science \textbf{359}, 301 (2018){]}. 
\end{abstract}
\maketitle
Over the past few years, a newly discovered phase of ultracold, dilute
quantum droplets has attracted increasingly attention in different
fields of physics \cite{Petrov2018,FerrierBarbut2019,Kartashov2019,Bottcher2020}.
In sharp contrast to other gas-like phases in containers, quantum
droplets are self-bound, liquid-like clusters of ten to hundred thousands
of atoms in free space, formed by the delicate balance between the
attractive mean-field force and repulsive force from quantum fluctuations
\cite{FerrierBarbut2016,Schmitt2016,Chomaz2016,Cabrera2018,Semeghini2018,Bottcher2019}.
A prototype theory of such quantum droplets was constructed by Petrov
in his seminal work \cite{Petrov2015} for a three-dimensional Bose-Bose
mixture with intraspecies repulsions and interspecies attractions,
characterized by the $s$-wave scattering lengths $a_{11}>0$, $a_{22}>0$,
and $a_{12}<0$, respectively. Using the conventional Bogoliubov theory
for Bose-Einstein condensates (BEC) \cite{Larsen1963} with modification
(referred to as Petrov's theory hereafter), Petrov showed that the
mechanical collapse at the condition $-a_{12}>a\equiv\sqrt{a_{11}a_{22}}$
anticipated from the mean-field picture can be stabilized by the first-order
Lee-Huang-Yang (LHY) correction due to quantum fluctuations \cite{LeeHuangYang1957}.
This surprising proposal has now been experimentally confirmed in
bosonic homonuculear $^{39}$K-$^{39}$K mixtures \cite{Cabrera2018,Semeghini2018,Cheiney2018,Ferioli2019}
and heteronuclear $^{41}$K-$^{87}$Rb mixtures \cite{DErrico2019}.
Petrov's theory is also generalized to different setups and configurations
\cite{Petrov2016,Cappellaro2017,Li2017,Cui2018,Jorgensen2018,Shi2019,Wang2020},
providing an important starting point to understand intriguing many-body
effects beyond mean-field. A lot of numerical studies beyond the LHY
correction have then been motivated, including numerically accurate
diffusion Monte Carlo (DMC) technique in various dimensions \cite{Petrov2016,Cikojevic2019,Parisi2019,Cikojevic2020}.

While Petrov's theory successfully captures the essential features
of quantum droplets, there is an annoying intrinsic inconsistency.
As the mean-field theory predicts a collapsing phase, one of the two
gapless Bogoliubov spectra necessarily gets softened and becomes complex
\cite{Petrov2015}. As a consequence, the related LHY term is then
ill-defined. To overcome this technical difficulty, Petrov took an
approximate LHY term on the verge of the collapse (i.e., at $\delta a=a+a_{12}=0$),
by assuming its weak dependence on $\delta a$ \cite{Petrov2015}.
This approximation was recently examined by DMC simulations \cite{Cikojevic2019}.
While there is a reasonable agreement in the overall energy functional,
the equilibrium density of quantum droplets calculated from DMC shows
a notable decrease in comparison with the prediction of Petrov's theory,
even when $\left|\delta a\right|$ is relatively small \cite{Cikojevic2019}.
A similar significant decrease in the critical number of quantum droplets
was also observed in the first experimental realization \cite{Cabrera2018},
which can not be fully accounted by Petrov's theory and remains to
be theoretically understood so far \cite{Cikojevic2020}.

The purpose of this work is to develop a \emph{consistent} theory
of quantum droplets without the loophole of an approximate LHY term.
Our key idea is that, in the presence of interspecies attractions,
two bosons in different species can form a bosonic pair, similar to
the well-known Cooper pair of two fermions with unlike spins in conventional
Bardeen--Cooper--Schrieffer (BCS) superconductors \cite{BCS1957}.
The generalization of the Bogoliubov theory with the inclusion of
the bosonic pairing then leads to two well-defined Bogoliubov spectra,
in which the previously softening mode in Petrov's theory now becomes
gapped, as a result of pairing. 

With this correct description of the ground state, we find unexpectedly
that, a rigorous treatment of the regularization of the contact interactions,
which is often overlooked for weakly interacting Bose gases, renormalizes
both the mean-field energy and the LHY correction. In comparison with
Petrov's theory, the mean-field energy is weakened by a factor of
$-a/a_{12}<1$ and the LHY term is approximately enlarged by a factor
of $(1+x^{2})/(2x)$, where $x\equiv(a_{11}/a_{22})^{1/4}$. As a
result, the equilibrium density of quantum droplets can decrease significantly,
already at the relatively small $\left|\delta a\right|\sim0.2a$,
in agreement with the recent DMC finding \cite{Cikojevic2019}. 

Our consistent theory opens the possibility of \emph{quantitatively}
describing self-bound quantum droplets with ultracold atoms towards
the strongly correlated regime, which could be termed as bosonic BEC-BCS
crossover. It can also be naturally generalized to take into account
the spatial inhomogeneity of the droplets, without the commonly-used
local density approximation or density functional theory \cite{Petrov2015,Cikojevic2019,Cikojevic2020}.
Thus it can provide an accurate description of collective oscillations
of this new quantum phase, which is of great interest in on-going
experiments \cite{Cabrera2018,Semeghini2018}. Our results may also
be useful to understand strongly interacting droplet phases in other
contexts, such as nanometer-sized clusters of helium atoms \cite{Stringari1987,Dalfovo1994,Laimer2019}
and electron-hole droplets in semiconductors \cite{AlmandHunter2014,Arp2019}.

\textit{Model Hamiltonian}. To be concrete, we consider a homonuclear
Bose-Bose mixture in three dimensions, described by the model Hamiltonian
$\mathscr{H}=\mathscr{H}_{0}+\mathscr{H}_{\textrm{int}}$ as 
\begin{eqnarray}
\mathscr{H}_{0} & = & \sum_{\mathbf{k},i=1,2}\left(\varepsilon_{\mathbf{k}}-\mu_{i}\right)\phi_{i\mathbf{k}}^{\dagger}\phi_{i\mathbf{k}},\label{eq: HamiCoul}\\
\mathscr{H}_{\textrm{int}} & = & \sum_{ij=1,2}\frac{g_{ij}}{2\mathcal{V}}\sum_{\mathbf{k}\mathbf{k}'\mathbf{q}}\phi_{i\mathbf{q}-\mathbf{k}}^{\dagger}\phi_{j\mathbf{k}}^{\dagger}\phi_{j\mathbf{q}-\mathbf{k}'}\phi_{i\mathbf{k}'},
\end{eqnarray}
where $\phi_{i\mathbf{k}}$ are the annihilation field operators of
the $i$-species bosons with same mass $m$ and dispersion $\varepsilon_{\mathbf{k}}\equiv\hbar^{2}\mathbf{k}^{2}/(2m)$,
$\mu_{i}$ are the chemical potentials to be fixed by the number of
atoms $n_{i}$, $\mathcal{V}$ is the volume and is taken to be unity
hereafter, and $g_{ij}$ are the bare intraspecies and interspecies
interaction strengths, which can be regularized using the $s$-wave
scattering length $a_{\ensuremath{ij}}$, i.e., 
\begin{equation}
\frac{1}{g_{ij}}=\frac{m}{4\pi\hbar^{2}a_{ij}}-\sum_{\mathbf{k}}\frac{m}{\hbar^{2}\mathbf{k}^{2}}.\label{eq:gR}
\end{equation}
Quantum droplets emerges once the repulsive intraspecies interactions
are less than the attractive interspecies interactions \cite{Petrov2015},
i.e., $\sqrt{a_{11}a_{22}}=a<-a_{12}$.

\textit{Petrov's theory.} We start by briefly reviewing Petrov's theory
of quantum droplets for equal intraspecies interactions $a_{11}=a_{22}=a$
and $n_{1}=n_{2}=n/2$. In this case, the energy per particle at zero
temperature predicted by the Bogoliubov theory is given by \cite{Larsen1963,Cikojevic2019},
\begin{equation}
\frac{E}{N}=\frac{\pi\hbar^{2}}{m}\left(a+a_{12}\right)n+\frac{32\sqrt{2\pi}}{15}\frac{\hbar^{2}a^{5/2}}{m}\mathcal{F}\left(\frac{a_{12}}{a}\right)n^{3/2},\label{eq:EnergyPetrov}
\end{equation}
where $\mathcal{F}(\alpha)\equiv(1+\alpha)^{5/2}+(1-\alpha)^{5/2}$
becomes complex in the droplet phase $a+a_{12}<0$. This is caused
by the imaginary sound velocity $c^{2}=2\pi\hbar^{2}(a+a_{12})n/m^{2}<0$,
signifying a collapse mean-field solution. To solve this issue, one
may approximate $\mathcal{F}(a_{12}/a)\simeq\mathcal{F}(1)=4\sqrt{2}$
\cite{Petrov2015}, despite the fact that $\textrm{Re}\mathcal{F}(\alpha)$
is a rapidly changing function. This approximation leads to an equilibrium
density \cite{Petrov2015,Cikojevic2019} 
\begin{equation}
n_{0}=\frac{25\pi}{16384}\left(1+\frac{a_{12}}{a}\right)^{2}a^{-3},
\end{equation}
 at which $E/N$ takes the minimum.

\textit{Bosonic pairing theory}. As a complex sound mode indicating
an unstable ground state, we would rather be interested in finding
the \emph{true} ground state with all positive excitation spectra.
This is particularly relevant in developing quantitatively reliable
theory of quantum droplets. Our key observation is that the attractive
interspecies interactions may induce a pairing of two bosons in different
species, analogous to their fermionic counterpart at the BEC-BCS crossover
\cite{BCS1957,Hu2006,Hu2007}. To verify this idea, we decouple the
interspecies interaction Hamiltonian using Hubbard--Stratonovich
transformation with a pairing field at the saddle-point level $\Delta=-g_{12}\sum_{\mathbf{k}}\left\langle \phi_{1\mathbf{k}}\phi_{2-\mathbf{k}}\right\rangle >0$
\cite{Hu2006}, which yields the terms $-\Delta^{2}/g_{12}-\Delta\sum_{\mathbf{k}}(\phi_{1\mathbf{k}}\phi_{2-\mathbf{k}}+\textrm{H.c.})$.

At zero temperature, we assume that the two bosonic fields condense
into the zero-momentum state with wave-function $\phi_{ic}\propto\sqrt{n_{i}}$.
At the leading order, the thermodynamic potential from condensates
takes the form,
\begin{equation}
\varOmega_{0}=-\frac{\Delta^{2}}{g_{12}}-2\Delta\phi_{1c}\phi_{2c}+\sum_{i=1,2}\left(-\mu_{i}\phi_{ic}^{2}+\frac{g_{ii}}{2}\phi_{ic}^{4}\right).
\end{equation}
By defining $C_{i}=g_{ii}\phi_{ic}^{2}$ and minimizing $\varOmega_{0}$
with respect to $\phi_{ic}$, we obtain $C_{1}=\mu_{1}+\Delta(\phi_{2c}/\phi_{1c})$,
$C_{2}=\mu_{2}+\Delta(\phi_{1c}/\phi_{2c})$ and $\varOmega_{0}=-\Delta^{2}/g_{12}-C_{1}^{2}/(2g_{11})-C_{2}^{2}/(2g_{22})$.
The next-order contribution to the thermodynamic potential comes from
Gaussian fluctuations around the condensates, described by the bilinear
Hamiltonian,
\begin{eqnarray}
\mathscr{H}_{\textrm{Bog}} & = & \sum_{i=1,2}\sum_{\mathbf{k\neq0}}\left[B_{i\mathbf{k}}\phi_{i\mathbf{k}}^{\dagger}\phi_{i\mathbf{k}}+\frac{C_{i}}{2}\left(\phi_{i\mathbf{k}}^{\dagger}\phi_{i\mathbf{-k}}^{\dagger}+\textrm{H.c.}\right)\right]\nonumber \\
 &  & -\sum_{\mathbf{k\neq0}}\Delta\left(\phi_{1\mathbf{k}}^{\dagger}\phi_{2\mathbf{k}}^{\dagger}+\textrm{H.c.}\right),
\end{eqnarray}
where $B_{i\mathbf{k}}\equiv\varepsilon_{\mathbf{k}}-\mu_{i}+2C_{i}$.
By diagonalizing $\mathscr{H}_{\textrm{Bog}}$, we obtain two Bogoliubov
spectra, $E_{\pm}^{2}(\mathbf{k})=[\mathcal{A}_{+}(\mathbf{k})-\Delta^{2}]\pm\{\mathcal{A}_{-}^{2}(\mathbf{k})+\Delta^{2}[(C_{1}+C_{2})^{2}-(B_{1\mathbf{k}}-B_{2\mathbf{k}})^{2}]\}^{1/2}$,
with $\mathcal{A}_{\pm}(\mathbf{k})\equiv[(B_{1\mathbf{k}}^{2}-C_{1}^{2})\pm(B_{2\mathbf{k}}^{2}-C_{2}^{2})]/2$.
Therefore, the fluctuation contribution to the thermodynamic potential
takes the form \cite{Salasnich2016,Hu2020},
\begin{equation}
\varOmega_{\textrm{LHY}}=\frac{1}{2}\sum_{\mathbf{k}}\left[E_{+}\left(\mathbf{k}\right)+E_{-}\left(\mathbf{k}\right)-B_{1\mathbf{k}}-B_{2\mathbf{k}}\right],
\end{equation}
which is formally ultraviolet divergent due to the use of contact
interactions. The divergence, however, can be exactly removed by regularizing
the bare interaction strengths using Eq. (\ref{eq:gR}). By adding
$\varOmega_{0}$ and $\varOmega_{\textrm{LHY}}$ together, we find
a finite sum, 
\begin{eqnarray}
\varOmega & = & -\frac{m}{4\pi\hbar^{2}}\left[\frac{C_{1}^{2}}{2a_{11}}+\frac{C_{2}^{2}}{2a_{22}}+\frac{\Delta^{2}}{a_{12}}\right]+\frac{1}{2}\sum_{\mathbf{k}}\left[E_{+}\left(\mathbf{k}\right)+\right.\nonumber \\
 &  & \left.+E_{-}\left(\mathbf{k}\right)-B_{1\mathbf{k}}-B_{2\mathbf{k}}+\frac{C_{1}^{2}+C_{2}^{2}+2\Delta^{2}}{\hbar^{2}\mathbf{k}^{2}/m}\right].
\end{eqnarray}
To determine the pairing parameter $\Delta=\Delta_{0}$, for given
chemical potentials $\mu_{i}$ we minimize the thermodynamic potential
$\varOmega$ with respect to $\Delta$. We note that $E_{-}(\mathbf{k}\rightarrow0)=0$
and hence the lower Bogoliubov branch is gapless. In contrast, the
upper Bogoliubov branch has a gap.

\begin{figure}[t]
\begin{centering}
\includegraphics[width=0.48\textwidth]{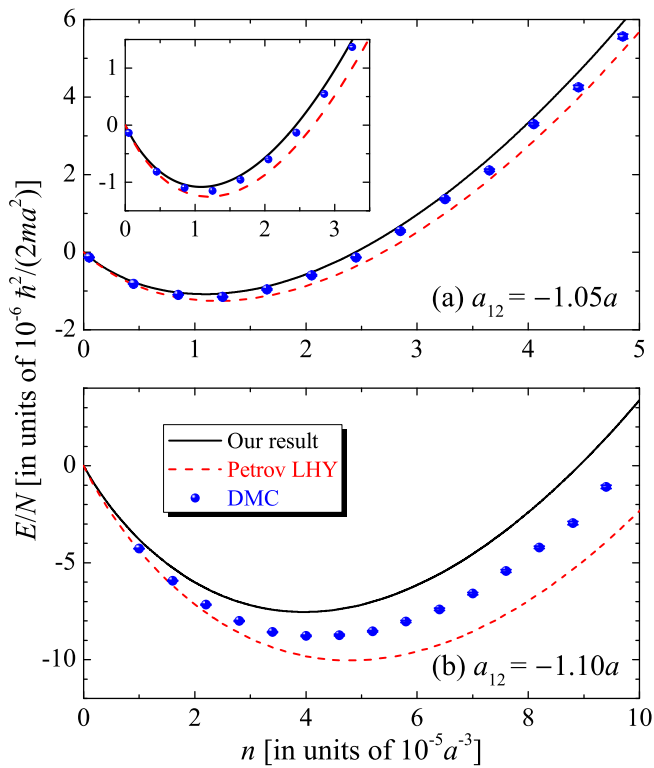}
\par\end{centering}
\caption{\label{fig1_energy} Energy per particle as a function of the density
at the interspecies interaction $a_{12}=-1.05a$ (a) and $a_{12}=-1.10a$
(b) and at the equal intraspecies interactions $a_{11}=a_{22}\equiv a$.
Our results (black solid line) are compared with Petrov's MF + LHY
prediction (red dashed line) \cite{Petrov2015} and the recent DMC
data (blue circles) \cite{Cikojevic2019}. The inset in (a) highlights
the comparison near the equilibrium density.}
\end{figure}

\textit{Equal intraspecies interactions}. To see this, let us first
focus on the idealized case of $a_{11}=a_{22}$, with which we take
$\mu_{1}=\mu_{2}=\mu$ and $\phi_{1c}=\phi_{2c}$, so that $C_{1}=C_{2}=\mu+\Delta>0$
and $B_{1\mathbf{k}}=B_{2\mathbf{k}}=\varepsilon_{\mathbf{k}}+\mu+2\Delta$.
The lower and upper Bogoliubov spectra then have the form, $E_{-}(\mathbf{k})=\sqrt{\varepsilon_{\mathbf{k}}(\varepsilon_{\mathbf{k}}+2\mu+4\Delta)}$
and $E_{+}(\mathbf{k})=\sqrt{(\varepsilon_{\mathbf{k}}+2\mu+2\Delta)(\varepsilon_{\mathbf{k}}+2\Delta)}$,
respectively. The upper Bogoliubov branch clearly shows an energy
gap $E_{\textrm{g}}=2\Delta\sqrt{1+\mu/\Delta}$. Hence, the unstable
branch in Petrov's theory is automatically removed with the introduction
of the bosonic pairing. This also implies that we obtain the true
ground state of quantum droplets.

We find that at $a_{11}=a_{22}$ the thermodynamic potential becomes
($C=\mu+\Delta$),
\begin{equation}
\varOmega=-\frac{m}{4\pi\hbar^{2}}\left[\frac{C^{2}}{a}+\frac{\Delta^{2}}{a_{12}}\right]+\frac{8m^{3/2}}{15\pi^{2}\hbar^{3}}C^{5/2}\mathcal{G}\left(\frac{\Delta}{C}\right),\label{eq:OmegaA11EqA22}
\end{equation}
where $h(\alpha)\equiv(15/4)\int_{0}^{\infty}dt\sqrt{t}[\sqrt{(t+1)(t+\alpha)}-(t+1/2+\alpha/2)+(1-\alpha)^{2}/(8t)]$
and $\mathcal{G}(\alpha)\equiv(1+\alpha)^{5/2}+h(\alpha)$ slightly
differs from $\mathcal{F}(\alpha)$ defined in Eq. (\ref{eq:EnergyPetrov}).
As discussed in detail in Supplemental Material \cite{SM}, for a
given chemical potential $\mu$ above a critical value $\mu_{c}<0$,
we typically find a minimum in $\varOmega(\Delta)$ located at the
pairing parameter $\Delta_{0}\neq0$. By calculating the density $n=-\partial\varOmega/\partial\mu$,
we then obtain the total energy per particle $E/N=\varOmega/n+\mu$
as a function of $n$, which clearly exhibits an absolute minimum
anticipated for quantum droplets. At $\mu<\mu_{c}$, $\Delta_{0}$
jumps to zero, indicating a first-order phase transition to a collapsing
state for sufficiently small densities \cite{CollapseNote}.

Numerically, we find $\left|\mu\right|\ll C,\Delta_{0}$, due to the
delicate balance in the first term in Eq. (\ref{eq:OmegaA11EqA22}).
As an excellent approximation, we neglect the $\mu$-dependence in
the second term and rewrite the regularized LHY thermodynamic potential
$\varOmega_{\textrm{LHY}}=[16(2m)^{3/2}/(15\pi^{2}\hbar^{3})]\Delta_{0}^{5/2}$.
The dominant $\mu$-dependence in the regularized $\varOmega_{0}$
then leads to $n\simeq m\Delta_{0}/(2\pi\hbar^{2}a)$. Replacing $\Delta_{0}$
by $n$, we obtain,
\begin{equation}
\frac{E}{N}=-\frac{\pi\hbar^{2}}{m}\left(a+\frac{a^{2}}{a_{12}}\right)n+\frac{256\sqrt{\pi}}{15}\frac{\hbar^{2}a^{5/2}}{m}n^{3/2}.\label{eq:EnergyA11EqA22}
\end{equation}
Compared with Eq. (\ref{eq:EnergyPetrov}), it is interesting to see
that the \emph{approximate} LHY term adopted by Petrov is reproduced
by our pairing theory and is actually \emph{exact} at the special
case of $a_{11}=a_{12}$. However, the mean-field energy, the first
term in Eq. (\ref{eq:EnergyA11EqA22}), is now changed by a factor
of $-a/a_{12}<1$. As a result, the equilibrium density becomes
\begin{equation}
n_{\textrm{eq}}=\frac{25\pi}{16384}\left(1+\frac{a}{a_{12}}\right)^{2}a^{-3}=\frac{a^{2}}{a_{12}^{2}}n_{0},
\end{equation}
and is reduced by a factor of $(a/a_{12})^{2}$, with respect to Petrov's
prediction $n_{0}$. 

\begin{figure}[t]
\begin{centering}
\includegraphics[width=0.48\textwidth]{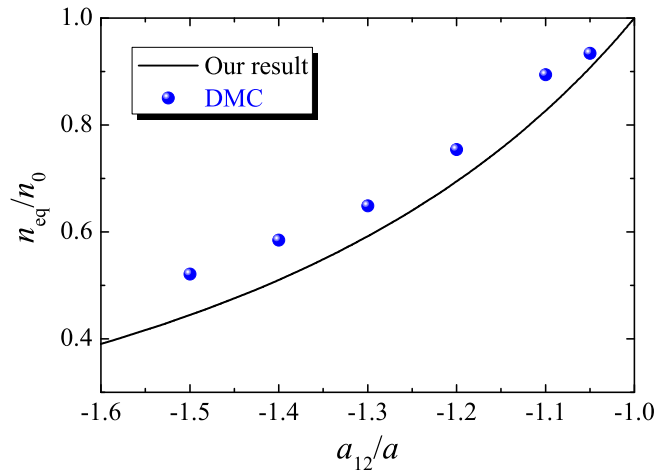} 
\par\end{centering}
\caption{\label{fig2_neq} Equilibrium density in units of $n_{0}=[25\pi/16384](1+a_{12}/a)^{2}a^{-3}$
(which is the equilibrium density predicted by Petrov's theory), as
a function of $a_{12}/a$. Our result (black solid line) agrees reasonably
well with the DMC data (blue circles). }
\end{figure}

In Fig. \ref{fig1_energy}, we show the density dependence of the
energy per particle given by Eq. (\ref{eq:EnergyA11EqA22}) (solid
line) and Eq. (\ref{eq:EnergyPetrov}) (dashed line) at the interspecies
interactions $a_{12}=-1.05a$ (a) and $a_{12}=-1.10a$ (b), and compare
them with the benchmark DMC results. We find a good agreement between
our result and the DMC data at smaller $\left|a_{12}\right|$ where
the gas parameter $na^{3}\sim10^{-5}$ is small, as exemplified in
the inset of Fig. \ref{fig1_energy}(a). At larger $\left|a_{12}\right|$
in (b), our result up-shifts from the DMC data, as the density becomes
larger. This is anticipated, as our pairing theory within the Bogoliubov
framework only predicts an \emph{upper} bound for the energy and the
higher-order three-body effect beyond LHY should come into a play
at density $na^{3}\sim5\times10^{-5}$ \cite{Wu1959}. In Fig. \ref{fig2_neq},
we report the ratio $n_{\textrm{eq}}/n_{0}$ as a function of $a_{12}/a$.
There is a reasonable agreement between our prediction and the DMC
data, although our theory becomes increasingly worse at larger $\left|a_{12}/a\right|$
due to the large equilibrium density.

\begin{figure}[t]
\begin{centering}
\includegraphics[width=0.48\textwidth]{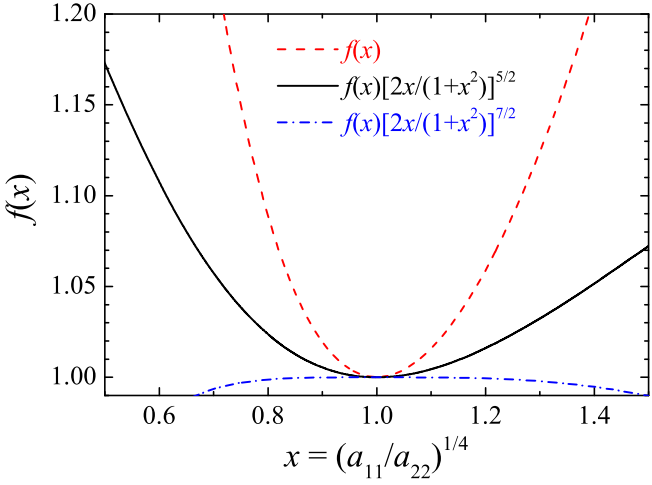}
\par\end{centering}
\caption{\label{fig3_fx} The functions $f(x)$ and $f(x)[2x/(1+x^{2})]^{5/2}$
as a function of $x=(a_{11}/a_{22})^{1/4}$. The latter measures the
enhancement of the LHY term in our pairing theory. As shown by the
blue dot-dashed line, $f(x)$ can be approximated by $[2x/(1+x^{2})]^{-7/2}$
at the interval $x\subseteq[0.8,1.25]$, within less than $0.1\%$
error in accuracy.}
\end{figure}

\begin{figure}[t]
\begin{centering}
\includegraphics[width=0.48\textwidth]{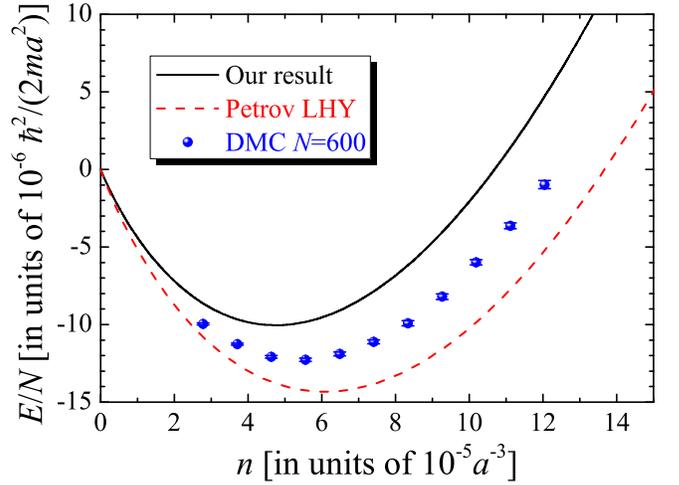}
\par\end{centering}
\caption{\label{fig4_energyK39} Energy per particle of a $^{39}$K-$^{39}$K
Bose mixture at the magnetic field $B=56.337$G, as a function of
the total density. Our result (black solid line) is compared with
Petrov's prediction (red dashed line) and the recent DMC data (blue
circles) at the number of atoms $N=600$. At this magnetic field,
$a_{11}=66.619a_{0}$, $a_{12}=-53.386a_{0}$ and $a_{22}=34.369a_{0}$.
We find that $a=(a_{11}a_{22})^{1/2}=47.85a_{0}$ and $x=(a_{11}/a_{22})^{1/4}\simeq1.180$.}
\end{figure}

\textit{Unequal intraspecies interactions}. Let us now consider the
realistic situation with unequal intraspecies interactions $a_{11}\neq a_{12}$.
It is useful to parametrize the imbalance in the densities by $x=\phi_{2c}/\phi_{1c}=\sqrt{n_{2}/n_{1}}$,
so that $C_{1}=\mu_{1}+\Delta x$ and $C_{2}=\mu_{2}+\Delta/x$. In
this case, $\left|\mu_{i}\right|\ll C_{i},\Delta_{0}$ near the equilibrium
density and it is still an excellent approximation to neglect the
$\mu$-dependence in $\varOmega_{\textrm{LHY}}.$ Therefore, we find
$\varOmega_{\textrm{LHY}}=[32\sqrt{2}m^{3/2}/(15\pi^{2}\hbar^{3})]\Delta^{5/2}f(x)$,
where the detailed expression of $f(x)$ is given in Supplemental
Material \cite{SM} and its value is shown in Fig. \ref{fig3_fx}.
In the interval of experimental interest, i.e., $x\subseteq[0.8,1.25]$,
to a great accuracy $f(x)\simeq[2x/(1+x^{2})]^{-7/2}.$ On the other
hand, the renormalized mean-field thermodynamic potential is given
by, $\varOmega_{0}=-[m/(8\pi\hbar^{2})][(\mu_{1}+\Delta x)^{2}/a_{11}+(\mu_{2}+\Delta/x)^{2}/a_{22}+2\Delta^{2}/a_{12}]$,
from which we obtain the densities, $n_{1}\simeq xm\Delta_{0}/(4\pi\hbar^{2}a_{11})$
and $n_{2}\simeq x^{-1}m\Delta_{0}/(4\pi\hbar^{2}a_{22})$. Hence,
\begin{equation}
x^{2}=\frac{n_{2}}{n_{1}}=\sqrt{\frac{a_{11}}{a_{22}}},
\end{equation}
as predicted by Petrov \cite{Petrov2015}. Replacing $\Delta_{0}$
again with the density $n=n_{1}+n_{2}$, we arrive at $E=E_{0}+E_{\textrm{LHY}}$,
\begin{eqnarray}
\frac{E_{0}}{N} & = & -\frac{\pi\hbar^{2}}{m}\left(a+\frac{a^{2}}{a_{12}}\right)\left[\frac{2x}{1+x^{2}}\right]^{2}n,\\
\frac{E_{\textrm{LHY}}}{N} & = & \frac{256\sqrt{\pi}}{15}\frac{\hbar^{2}a^{5/2}}{m}\left[\frac{2x}{1+x^{2}}\right]^{5/2}f\left(x\right)n^{3/2}.
\end{eqnarray}
Compared with Petrov's energy at $a_{11}\neq a_{22}$ \cite{Petrov2015},
we find that, in addition to the reduction in the mean-field energy
as in Eq. (\ref{eq:EnergyA11EqA22}), the LHY energy is enhanced by
a factor of $[2x/(1+x^{2})]^{5/2}f(x)\simeq(1+x^{2})/(2x)$. Therefore,
the equilibrium density
\begin{equation}
\frac{n_{\textrm{eq}}}{n_{0}}=\frac{a^{2}}{a_{12}^{2}}\left[\frac{1+x^{2}}{2x}\right]^{5}\frac{1}{f^{2}\left(x\right)}\simeq\frac{a^{2}}{a_{12}^{2}}\left[\frac{2x}{1+x^{2}}\right]^{2}
\end{equation}
 decreases further at $a_{11}\neq a_{22}$ compared to Petrov's prediction.
In Fig. \ref{fig4_energyK39}, we present the density dependence of
the energy per particle for a $^{39}$K Bose-Bose mixture at the magnetic
field $B=56.337$G. Our result is compared with the latest DMC data
with $N=600$ particles \cite{Cikojevic2020}, as well as Petrov's
prediction. The overall agreement with DMC data is reasonable, considering
the possible three-body effect beyond LHY at the moderately large
density \cite{Wu1959} and the finite-size effect at $N=600$ that
may slightly down-shift the DMC energy \cite{Cikojevic2019}. 

\textit{Experimental relevance}. Our observation of a reduced equilibrium
density in the pairing theory could be related to the smaller-than-expected
critical number of atoms found in the first experimental realization
of quantum droplets \cite{Cabrera2018}. However, for a quantitative
comparison, there are several important issues needed to take into
account. First, the effective range of interactions of the $^{39}$K-$^{39}$K
mixture is fairly large for both intraspecies and interspecies interactions
(i.e., $nr_{e}^{3}\sim0.1-1.0$), which significantly decreases the
energy functional \cite{Cikojevic2020}. Second, the three-body effect
may also play an important role at the gas parameter $na^{3}\sim10^{-4}$
\cite{Wu1959}. At last, an external harmonic trap may turn the experimental
setup into an effectively quasi-two-dimensional system \cite{Cabrera2018}.
These facts will be accounted for in our future studies. Furthermore,
in analogy to the conventional BCS superfluid, we anticipate that
pair fluctuations give rise to the gapless collective excitations
associated with the $U(1)$ symmetry breaking of the pairing field,
which can be potentially observed by applying Bragg spectroscopy in
future experiments \cite{Stenger1999}.

\textit{Conclusions}. We have developed a consistent theory of quantum
droplets and have refined the ground-breaking idea by Petrov that
the mean-field collapse can be prevented by quantum fluctuations.
Our correct construction of a pairing ground state paves the way to
investigate the bosonic BEC-BCS crossover and serves an ideal starting
point to explore the finite temperature effect and collective many-body
behavior of ultracold, ultradilute quantum droplets.
\begin{acknowledgments}
We are grateful to Tao Shi for simulating discussions, Viktor Cikojevi\'{c}
for sharing their DMC data, and Hui Deng for informing us the work
on electron-hole droplets. This research was supported by the Australian
Research Council's (ARC) Discovery Program, Grant No. DP170104008
(H.H.) and Grant No. DP180102018 (X.-J.L).
\end{acknowledgments}

\appendix

\begin{widetext}

\section{The Bogoliubov theory with bosonic pairing}

We use the conventional Bogoliubov theory with pairing to solve a
Bose-Bose mixture in the presence of attractive interspecies interaction.
The Hamiltonian density of the mixture in real space takes the form,
\begin{equation}
\mathscr{H}\left(\mathbf{x}\right)=\phi_{1}^{\dagger}\left(-\frac{\hbar^{2}\nabla^{2}}{2m}-\mu_{1}\right)\phi_{1}+\phi_{2}^{\dagger}\left(-\frac{\hbar^{2}\nabla^{2}}{2m}-\mu_{2}\right)\phi_{2}+\frac{g_{11}}{2}\phi_{1}^{\dagger}\phi_{1}^{\dagger}\phi_{1}\phi_{1}+\frac{g_{22}}{2}\phi_{2}^{\dagger}\phi_{2}^{\dagger}\phi_{2}\phi_{2}+g_{12}\phi_{1}^{\dagger}\phi_{2}^{\dagger}\phi_{2}\phi_{1},
\end{equation}
where $\phi_{i}(\mathbf{x})$ ($i=1,2$) is the annihilation field
operator of the $i$-species bosons and $\mu_{i}$ is the chemical
potential. The bare interaction strengths $g_{ij}$ are to be replaced
by the corresponding $s$-wave scattering length $a_{ij}$, via,
\begin{equation}
\frac{1}{g_{ij}}=\frac{m}{4\pi\hbar^{2}a_{ij}}-\sum_{k}\frac{m}{\hbar^{2}\mathbf{k}^{2}}.
\end{equation}
We are interested in calculating the thermodynamic potential $\varOmega$
from the partition function, using the path-integral formalism, $\mathcal{Z}=\int\mathcal{D}[\phi_{1},\phi_{2}]e^{-\mathcal{S}},$
where the action is given by,
\begin{equation}
\mathcal{S}=\int dx\left[\bar{\phi}_{1}\left(x\right)\partial_{\tau}\phi_{1}\left(x\right)+\bar{\phi}_{2}\left(x\right)\partial_{\tau}\phi_{2}\left(x\right)+\mathscr{H}\left(x\right)\right].
\end{equation}
Here, we have used the standard notations $x\equiv(\mathbf{x},\tau)$
and $\int dx\equiv\int d\mathbf{x}\int_{0}^{\beta}d\tau$, and $\beta\equiv1/(k_{B}T)$.

Due to the attractive interspecies interaction ($g_{12}<0$), we may
anticipate the pairing between different species. Therefore, we use
the Hubbard--Stratonovich (HS) transformation to decouple the last
term in the Hamiltonian density,
\begin{equation}
e^{-g_{12}\int dx\bar{\phi}_{1}\bar{\phi}_{2}\phi_{2}\phi_{1}}=\int\mathcal{D}\left[\Delta\left(x\right)\right]\exp\left\{ \int dx\left[\frac{\left|\Delta\left(x\right)\right|^{2}}{g_{12}}+\left(\bar{\Delta}\phi_{2}\phi_{1}+\bar{\phi}_{1}\bar{\phi}_{2}\Delta\right)\right]\right\} .
\end{equation}
The action then takes the form,
\begin{equation}
\mathcal{S}=\int dx\left\{ -\frac{\left|\Delta\left(x\right)\right|^{2}}{g_{12}}-\left(\bar{\Delta}\phi_{2}\phi_{1}+\bar{\phi}_{1}\bar{\phi}_{2}\Delta\right)+\sum_{i=1,2}\left[\bar{\phi}_{i}\left(\partial_{\tau}-\frac{\hbar^{2}\nabla^{2}}{2m}-\mu_{i}\right)\phi_{i}+\frac{g_{ii}}{2}\bar{\phi}_{i}\bar{\phi}_{i}\phi_{i}\phi_{i}\right]\right\} .
\end{equation}
For the pairing field $\Delta(x)$, it suffices to take a \emph{uniform}
saddle-point solution $\Delta(x)=\Delta>0$. At the same level of
approximation, we assume the bosons condensate into the zero-momentum
states, i.e., 
\begin{eqnarray}
\phi_{i}\left(x\right) & = & \phi_{ic}+\delta\phi_{i}\left(x\right),
\end{eqnarray}
with a real positive $\phi_{ic}>0$, and we approximate the intraspecies
interaction terms (i.e., within the Bogoliubov approximation),
\begin{equation}
\frac{g_{ii}}{2}\bar{\phi}_{i}\bar{\phi}_{i}\phi_{i}\phi_{i}\simeq\frac{g_{ii}}{2}\phi_{ic}^{4}+2g_{ii}\phi_{ic}^{2}\delta\bar{\phi}_{i}\delta\phi_{i}+\frac{g_{ii}\phi_{ic}^{2}}{2}\left(\delta\bar{\phi}_{i}\delta\bar{\phi}_{i}+\delta\phi_{i}\delta\phi_{i}\right).
\end{equation}
As a result, we find that $\mathcal{S}=\mathcal{S}_{0}+\mathcal{S}_{B}$,
where,
\begin{eqnarray}
S_{0} & = & \beta\mathcal{V}\left[\sum_{i=1,2}\left(-\mu_{i}\phi_{ic}^{2}+\frac{g_{ii}}{2}\phi_{ic}^{4}\right)-\frac{\Delta^{2}}{g_{12}}-2\Delta\phi_{1c}\phi_{2c}\right],\\
\mathcal{S}_{B} & = & \int dx\left\{ \sum_{i=1,2}\left[\delta\bar{\phi}_{i}\left(\partial_{\tau}-\frac{\hbar^{2}\nabla^{2}}{2m}-\mu_{i}+2g_{ii}\phi_{ic}^{2}\right)\delta\phi_{i}+\frac{g_{ii}\phi_{ic}^{2}}{2}\left(\delta\bar{\phi}_{i}\delta\bar{\phi}_{i}+\delta\phi_{i}\delta\phi_{i}\right)\right]-\Delta\left(\delta\phi_{2}\delta\phi_{1}+\delta\bar{\phi}_{1}\delta\bar{\phi}_{2}\right)\right\} .
\end{eqnarray}
By introducing the notations $C_{i}=g_{ii}\phi_{ic}^{2}$ and a Nambu
spinor $\Phi(x)=[\delta\phi_{1}(x),\delta\bar{\phi}_{1}(x),\delta\phi_{2}(x),\delta\bar{\phi}_{2}(x)]^{T}$,
we may rewrite $\mathcal{S}_{B}$ into a compact form,
\begin{equation}
\mathcal{S}_{B}=\int dxdx'\bar{\Phi}\left(x\right)\left[-\mathscr{D}^{-1}\left(x,x'\right)\right]\Phi\left(x'\right),
\end{equation}
where the inverse Green function of bosons is given by, 
\begin{equation}
-\mathscr{D}^{-1}=\left[\begin{array}{cccc}
\partial_{\tau}-\frac{\hbar^{2}\nabla^{2}}{2m}-\mu_{1}+2C_{1} & C_{1} & 0 & -\Delta\\
C_{1} & -\partial_{\tau}-\frac{\hbar^{2}\nabla^{2}}{2m}-\mu_{1}+2C_{1} & -\Delta & 0\\
0 & -\Delta & \partial_{\tau}-\frac{\hbar^{2}\nabla^{2}}{2m}-\mu_{2}+2C_{2} & C_{2}\\
-\Delta & 0 & C_{2} & -\partial_{\tau}-\frac{\hbar^{2}\nabla^{2}}{2m}-\mu_{2}+2C_{2}
\end{array}\right].
\end{equation}
We do not explicitly show the delta function $\delta\left(x-x'\right)$
in $\mathscr{D}^{-1}(x,x')$. By taking a Fourier transform, then,
in momentum space the bosonic Green function takes the form (after
transforming $\partial_{\tau}\rightarrow-i\omega_{m}$ and taking
the bosonic Matasubara frequencies, $i\omega_{m}\rightarrow\omega$),

\begin{equation}
\mathscr{D}^{-1}\left(\mathbf{k},\omega\right)=\left[\begin{array}{cccc}
\omega-B_{1\mathbf{k}} & -C_{1} & 0 & \Delta\\
-C_{1} & -\omega-B_{1\mathbf{k}} & \Delta & 0\\
0 & \Delta & \omega-B_{2\mathbf{k}} & -C_{2}\\
\Delta & 0 & -C_{2} & -\omega-B_{2\mathbf{k}}
\end{array}\right],
\end{equation}
where we have defined,
\begin{equation}
B_{i\mathbf{k}}\equiv\frac{\hbar^{2}\mathbf{k}^{2}}{2m}-\mu_{i}+2C_{i}=\varepsilon_{\mathbf{k}}-\mu_{i}+2C_{i}.
\end{equation}
By solving the poles of the bosonic Green function, i.e., $\det[\mathscr{D}^{-1}(\mathbf{k},\omega\rightarrow E(\mathbf{k}))]=0$,
or more explicitly, 
\begin{equation}
\omega^{4}-\omega^{2}\left[\left(B_{1\mathbf{k}}^{2}-C_{1}^{2}\right)+\left(B_{2\mathbf{k}}^{2}-C_{2}^{2}\right)-2\Delta^{2}\right]+\left[\left(B_{1\mathbf{k}}^{2}-C_{1}^{2}\right)\left(B_{2\mathbf{k}}^{2}-C_{2}^{2}\right)-2\left(B_{1\mathbf{k}}B_{2\mathbf{k}}+C_{1}C_{2}\right)\Delta^{2}+\Delta^{4}\right]=0,\label{eq:Det}
\end{equation}
we obtain the two Bogoliubov spectra, 
\begin{equation}
E_{\pm}^{2}\left(\mathbf{k}\right)=\left[\mathcal{A}_{+}\left(\mathbf{k}\right)-\Delta^{2}\right]\pm\sqrt{\mathcal{A}_{-}^{2}\left(\mathbf{k}\right)+\Delta^{2}\left[\left(C_{1}+C_{2}\right)^{2}-\left(B_{1\mathbf{k}}-B_{2\mathbf{k}}\right)^{2}\right]},
\end{equation}
with 
\begin{equation}
\mathcal{A}_{\pm}\left(\mathbf{k}\right)=\frac{\left(B_{1\mathbf{k}}^{2}-C_{1}^{2}\right)\pm\left(B_{2\mathbf{k}}^{2}-C_{2}^{2}\right)}{2}.
\end{equation}

\subsection{Thermodynamic potential from the condensate}

From the condensate contribution $\mathcal{S}_{0}$, we write down
the corresponding thermodynamic potential at the tree level,
\begin{equation}
\varOmega_{0}=-\frac{\Delta^{2}}{g_{12}}-2\Delta\phi_{1c}\phi_{2c}+\left(-\mu_{1}\phi_{1c}^{2}+\frac{g_{11}}{2}\phi_{1c}^{4}\right)+\left(-\mu_{2}\phi_{2c}^{2}+\frac{g_{22}}{2}\phi_{2c}^{4}\right).
\end{equation}
By taking the derivative of $\varOmega_{0}$ with respect to $\phi_{1c}$
and $\phi_{2c}$, we obtain,
\begin{eqnarray}
-\mu_{1}\phi_{1c}+g_{11}\phi_{1c}^{3}-\Delta\phi_{2c} & = & 0,\\
-\mu_{2}\phi_{2c}+g_{22}\phi_{2c}^{3}-\Delta\phi_{1c} & = & 0.
\end{eqnarray}
Therefore, we have,
\begin{eqnarray}
-\mu_{1}+C_{1}=B_{1\mathbf{k}=0}-C_{1} & = & \Delta\frac{\phi_{2c}}{\phi_{1c}},\\
-\mu_{2}+C_{2}=B_{2\mathbf{k}=0}-C_{2} & = & \Delta\frac{\phi_{1c}}{\phi_{2c}}.
\end{eqnarray}
It is easy to see that, 
\begin{equation}
\left(B_{1\mathbf{k}=0}-C_{1}\right)\left(B_{2\mathbf{k}=0}-C_{2}\right)=\Delta^{2}.
\end{equation}
As the last term in Eq. (\ref{eq:Det}) can be rewritten as,
\begin{equation}
\left[\left(B_{1\mathbf{k}}-C_{1}\right)\left(B_{2\mathbf{k}}-C_{2}\right)-\Delta^{2}\right]\left[\left(B_{1\mathbf{k}}+C_{1}\right)\left(B_{2\mathbf{k}}+C_{2}\right)-\Delta^{2}\right],
\end{equation}
the term is zero at $\mathbf{k}=0$. Thus, we confirm that at least
one of the two Bogoliubov spectra is gapless. This is anticipated
from the $U(1)$ symmetry breaking of the system. On the other hand,
it is also straightforward to confirm that,
\begin{equation}
\varOmega_{0}=-\frac{\Delta^{2}}{g_{12}}-\frac{C_{1}^{2}}{2g_{11}}-\frac{C_{2}^{2}}{2g_{22}}=-\frac{m}{4\pi\hbar^{2}}\left[\frac{\Delta^{2}}{a_{12}}+\frac{C_{1}^{2}}{2a_{11}}+\frac{C_{2}^{2}}{2a_{22}}\right]+\frac{1}{2}\sum_{\mathbf{k}}\frac{C_{1}^{2}+C_{2}^{2}+2\Delta^{2}}{\hbar^{2}\mathbf{k}^{2}/m},
\end{equation}
where in the last step, we have replaced the bare interaction strengths
by using the $s$-wave scattering lengths.

\subsection{LHY thermodynamic potential}

The LHY thermodynamic potential at the one-loop level can obtained
from $\mathcal{S}_{B}$ \cite{Salasnich2016,Hu2020},
\begin{equation}
\varOmega_{\textrm{LHY}}=\frac{k_{B}T}{2}\sum_{\mathbf{q},i\omega_{m}}\ln\det\left[\mathscr{-D}^{-1}\left(\mathbf{q},i\omega_{m}\right)\right]e^{i\omega_{m}0^{+}}=\frac{1}{2}\sum_{\mathbf{k}}\left[E_{+}\left(\mathbf{k}\right)+E_{-}\left(\mathbf{k}\right)-B_{1\mathbf{k}}-B_{2\mathbf{k}}\right].
\end{equation}
By putting together $\varOmega_{0}$ and $\varOmega_{\textrm{LHY}},$we
obtain the thermodynamic potential within the Bogoliubov approximation,
\begin{equation}
\varOmega=-\frac{m}{4\pi\hbar^{2}}\left[\frac{\Delta^{2}}{a_{12}}+\frac{C_{1}^{2}}{2a_{11}}+\frac{C_{2}^{2}}{2a_{22}}\right]+\frac{1}{2}\sum_{\mathbf{k}}\left[E_{+}\left(\mathbf{k}\right)+E_{-}\left(\mathbf{k}\right)-B_{1\mathbf{k}}-B_{2\mathbf{k}}+\frac{C_{1}^{2}+C_{2}^{2}+2\Delta^{2}}{\hbar^{2}\mathbf{k}^{2}/m}\right].
\end{equation}

\section{Equal intraspecies interactions}

In this case, $C_{1}=C_{2}=C=\mu+\Delta>0$ and $B_{1\mathbf{k}}=B_{2\mathbf{k}}=\varepsilon_{\mathbf{k}}+\mu+2\Delta$,
and the thermodynamic potential is given by,
\begin{equation}
\varOmega=-\frac{m}{4\pi\hbar^{2}}\left[\frac{C^{2}}{a}+\frac{\Delta^{2}}{a_{12}}\right]+\frac{1}{2}\sum_{\mathbf{k}}\left[E_{+}\left(\mathbf{k}\right)+E_{-}\left(\mathbf{k}\right)-2\left(\varepsilon_{\mathbf{k}}+C+\Delta\right)+\frac{2\left(C^{2}+\Delta^{2}\right)}{\hbar^{2}\mathbf{k}^{2}/m}\right],
\end{equation}
where $\varepsilon_{\mathbf{k}}\equiv\hbar^{2}k^{2}/(2m)$ and two
Bogoliubov spectra are, 
\begin{eqnarray}
E_{-}(\mathbf{k}) & = & \sqrt{\varepsilon_{\mathbf{k}}\left(\varepsilon_{\mathbf{k}}+2C+2\Delta\right)},\\
E_{+}(\mathbf{k}) & = & \sqrt{\left(\varepsilon_{\mathbf{k}}+2C\right)\left(\varepsilon_{\mathbf{k}}+2\Delta\right)}.
\end{eqnarray}
The integral can be separated into two parts $\mathcal{I}_{+}$ and
$\mathcal{I}_{-}$, where 
\begin{equation}
\mathcal{I}_{\pm}=\frac{1}{2}\sum_{\mathbf{k}}\left[E_{\pm}\left(\mathbf{k}\right)-\left(\varepsilon_{\mathbf{k}}+C+\Delta\right)+\frac{\left(C\pm\Delta\right)^{2}}{\hbar^{2}\mathbf{k}^{2}/m}\right].
\end{equation}
By introducing a new variable $t\equiv[\hbar^{2}k^{2}/(2m)]/[2(C+\Delta)]$,
it is easy to see that,
\begin{equation}
\mathcal{I}_{-}=\frac{1}{8\pi^{2}}\left(\frac{2m}{\hbar^{2}}\right)^{3/2}\left[2\left(C+\Delta\right)\right]^{5/2}\int_{0}^{\infty}dt\sqrt{t}\left[\sqrt{t\left(t+1\right)}-\left(t+\frac{1}{2}\right)+\frac{1}{8t}\right]=\frac{8m^{3/2}}{15\pi^{2}\hbar^{3}}C^{5/2}\left(1+\frac{\Delta}{C}\right)^{5/2}.
\end{equation}
To calculate $\mathcal{I}_{+}$, we instead introduce $t\equiv[\hbar^{2}k^{2}/(2m)]/(2C)$
and $\alpha\equiv\Delta/C$, which leads to,

\begin{equation}
\mathcal{I}_{+}=\frac{1}{8\pi^{2}}\left(\frac{2m}{\hbar^{2}}\right)^{3/2}\left[2C\right]^{5/2}\int_{0}^{\infty}dt\sqrt{t}\left[\sqrt{\left(t+1\right)\left(t+\alpha\right)}-\left(t+\frac{1+\alpha}{2}\right)+\frac{\left(1-\alpha\right)^{2}}{8t}\right]\equiv\frac{8m^{3/2}}{15\pi^{2}\hbar^{3}}C^{5/2}h\left(\alpha\right).
\end{equation}
By adding up $\mathcal{I}_{+}$ and $\mathcal{I}_{-}$, we obtain
the LHY term,
\begin{equation}
\varOmega_{\textrm{LHY}}=\mathcal{I}_{+}+\mathcal{I}_{-}=\frac{8m^{3/2}}{15\pi^{2}\hbar^{3}}\left(\mu+\Delta\right)^{5/2}\mathcal{G}\left(\frac{\Delta}{\mu+\Delta}\right),
\end{equation}
where $\mathcal{G}(\alpha)\equiv(1+\alpha)^{5/2}+h\left(\alpha\right)$.
It is easy to obtain that $h(0)=1$, $h(1)=0$, $h'(1)=0$ and $h''(1)=15\pi/16$.
Compared with the function $\mathcal{F}(\alpha)\equiv(1+\alpha)^{5/2}+(1-\alpha)^{5/2}$
for the LHY energy in the Bogoliubov theory of a Bose-Bose mixture,
we find the role of $(1-\alpha)^{5/2}$, which is not well-defined
for $\alpha>1$, is now replaced by a new function $h(\alpha)$. In
Fig. \ref{figSM1_hA}, we show the function $h(\alpha)$. It is larger
than $(1-\alpha)^{5/2}$ in the interval $\alpha\subseteq[0,1]$.

\begin{figure}[t]
\begin{centering}
\includegraphics[width=0.5\textwidth]{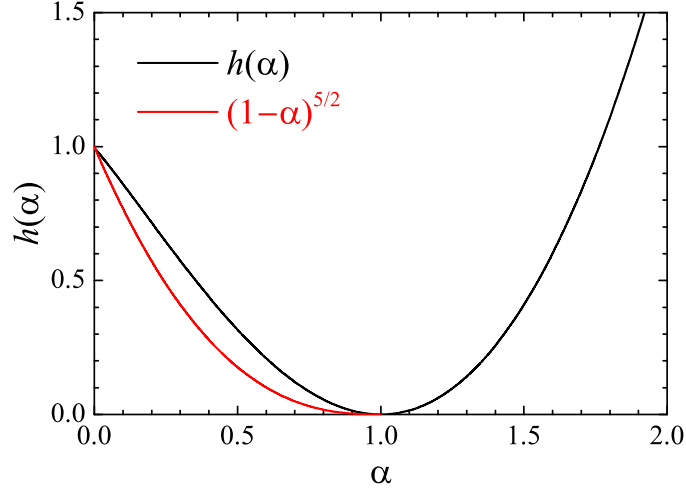}
\par\end{centering}
\caption{\label{figSM1_hA} The function $h(\alpha)$ and its comparison to
$(1-\alpha)^{5/2}$.}
\end{figure}

\begin{figure}[t]
\begin{centering}
\includegraphics[width=0.5\textwidth]{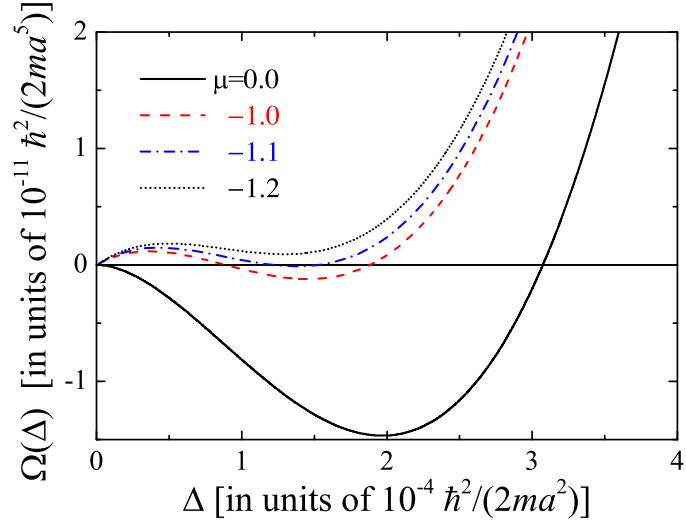}
\par\end{centering}
\caption{\label{figSM2_omega} Thermodynamic potential $\varOmega$, in units
of $10^{-11}\hbar^{2}/(2ma^{5})$, as a function of the pairing parameter
$\Delta$, at different chemical potentials $\mu=0$ (black solid
line), $-1.0$ (red dashed line), $-1.1$ (blue dot-dashed line),
and $-1.2$ (black dotted line), and at $a_{12}=-1.05a$. $\Delta$
and $\mu$ are measured in units of $10^{-4}\hbar^{2}/(2ma^{2})$
and $10^{-6}\hbar^{2}/(2ma^{2})$, respectively. The critical chemical
potential is about $\mu_{c}\simeq-1.1\times10^{-6}\hbar^{2}/(2ma^{2})$.}
\end{figure}

Let us now consider the total thermodynamic potential,

\begin{equation}
\varOmega=-\frac{m}{4\pi\hbar^{2}}\left[\frac{\left(\mu+\Delta\right)^{2}}{a}+\frac{\Delta^{2}}{a_{12}}\right]+\frac{8m^{3/2}}{15\pi^{2}\hbar^{3}}\left(\mu+\Delta\right)^{5/2}\mathcal{G}\left(\frac{\Delta}{\mu+\Delta}\right).
\end{equation}
For a given chemical potential $\mu$, we need to the minimize $\varOmega$
to determine the pairing order parameter $\Delta_{0}$, and then calculate
the total density of the system, i.e., $n=-\partial\varOmega/\partial\mu$.
In Fig. \ref{figSM2_omega}, we show the thermodynamic potential $\varOmega$
as a function of $\Delta$, at four different chemical potentials
$\mu=0$, $-1.0$, $-1.1$ , and $-1.2$, which is measured in units
of $10^{-4}\hbar^{2}/(2ma^{2})$, and at $a_{12}=-1.05a$. For the
chemical potential above a critical value, i.e., $\mu_{c}\simeq-1.1\times10^{-4}\hbar^{2}/(2ma^{2})$,
we typically find a global minimum in the thermodynamic potential
at $\Delta_{0}\neq0$. For $\mu<\mu_{c}$, it turns into a local minimum
and the thermodynamic potential takes the global minimum at $\Delta_{0}=0$.
The change of the global minimum position $\Delta_{0}$ is not continuous
at $\mu=\mu_{c}$, indicating a first-order quantum phase transition
into a collapsing phase.

\begin{figure}[t]
\begin{centering}
\includegraphics[width=0.5\textwidth]{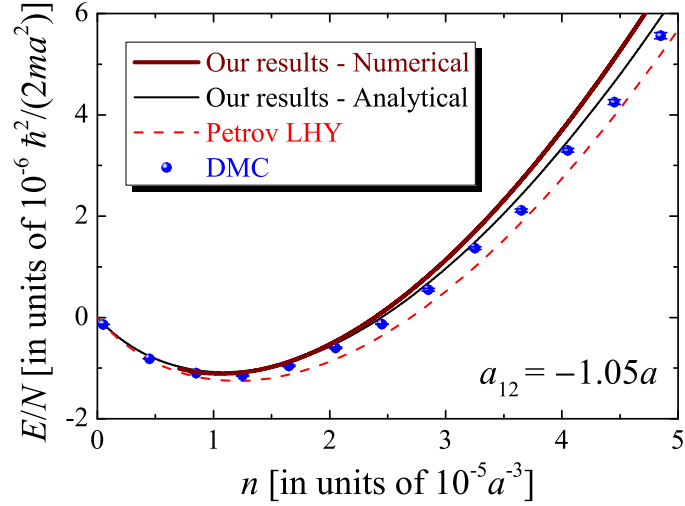}
\par\end{centering}
\caption{\label{figSM3_energy} Energy per particle as a function of the density
at the interspecies interaction $a_{12}=-1.05a$ and at the equal
intraspecies interactions $a_{11}=a_{22}\equiv a$. Our analytic result
(black solid line) is compared with Petrov's MF + LHY prediction (red
dashed line) \cite{Petrov2015} and the recent DMC data (blue circles)
\cite{Cikojevic2019}. Our full numerical result is shown by the brown
solid thick line.}
\end{figure}

\begin{figure}[t]
\begin{centering}
\includegraphics[width=0.5\textwidth]{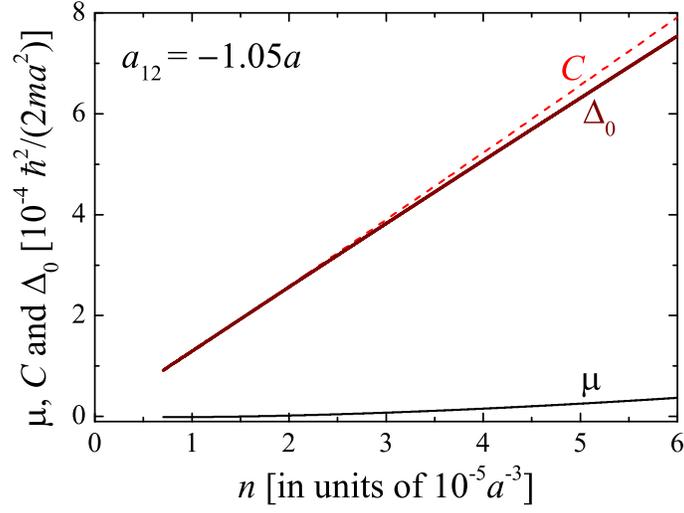}
\par\end{centering}
\caption{\label{figSM4_mu} Chemical potential $\mu$, the parameter $C$ and
the pairing gap $\Delta_{0}$, in units of $10^{-4}\hbar^{2}/(2ma^{2})$,
as a function of the total density $n$ (in units of $10^{-5}a^{-3}$)
at $a_{12}=-1.05a$.}
\end{figure}

For nonzero $\Delta_{0}\neq0$, we obtain $\varOmega(\mu,\Delta_{0})$
and calculate $n=-\partial\varOmega(\mu,\Delta_{0})/\partial\mu$.
Hence, we numerically obtain the total energy $E=\varOmega+\mu n$.
This energy is shown in Fig. \ref{figSM3_energy} by using a brown
thick solid line, as a function of the total density. It turns out
that numerically the chemical potential is much smaller than either
the parameter $C$ or the pairing gap $\Delta$ , as can be clearly
seen from Fig. \ref{figSM4_mu}. This is easy to understand from the
$\Delta$-dependence in $\varOmega_{0}$ and $\varOmega_{\textrm{LHY}}.$
We note that two terms in $\varOmega_{0}$ are large and have opposite
sign. Each of them (i.e., absolute value) is much larger than $\varOmega_{\textrm{LHY}}$.
Therefore, when we minimize $\varOmega$ with respect to $\Delta$,
we only need to minimize $\Omega_{0}$. This leads to the condition,
\begin{equation}
\frac{\mu+\Delta}{a}+\frac{\Delta}{a_{12}}\simeq0.
\end{equation}
Therefore, we find that,
\begin{equation}
\mu\simeq-\left(1+\frac{a}{a_{12}}\right)\Delta\ll\Delta,C
\end{equation}
as a result of $a_{12}\sim-a$.

Due to the smallness of $\left|\mu\right|$, it is reasonable to neglect
the $\mu$-dependence in $\varOmega_{\textrm{LHY}}$ and the term
$\mu^{2}$ in $\varOmega_{0}$. Therefore, we obtain,
\begin{equation}
\varOmega\simeq-\frac{m}{4\pi\hbar^{2}}\left[\frac{2\mu\Delta}{a}+\frac{\Delta^{2}}{a}+\frac{\Delta^{2}}{a_{12}}\right]+\frac{32\sqrt{2}m^{3/2}}{15\pi^{2}\hbar^{3}}\Delta^{5/2}.
\end{equation}
By taking the derivative with respect to $\mu$, we obtain 
\begin{equation}
n=-\frac{\partial\varOmega}{\partial\mu}\simeq\frac{m}{2\pi\hbar^{2}a}\Delta_{0},
\end{equation}
where we determine $\varOmega$ at the saddle point $\Delta=\Delta_{0}$.
Replacing the pairing parameter $\Delta_{0}$ by the density $n$,
we finally arrive at (the volume $\mathcal{V}=1$),
\begin{equation}
\frac{E}{N}=\frac{\varOmega}{n}+\mu=-\frac{\pi\hbar^{2}}{m}\left(a+\frac{a^{2}}{a_{12}}\right)n+\frac{256\sqrt{\pi}}{15}\frac{\hbar^{2}a^{5/2}}{m}n^{3/2}.
\end{equation}
In Fig. \ref{figSM3_energy}, this analytic result is shown by the
black solid line. We find an excellent agreement near the equilibrium
density between the analytic result and the full numerical result
for the energy per particle. However, for the density $n>3\times10^{-5}a^{-3}$,
the difference starts to become visible. This is not a serious problem,
as our perturbative treatment within in the Bogoliubov theory is anticipated
to become worse at similar densities. Thus, it is useless to quantify
the difference between the analytic and numerical results. 

\section{Unequal intraspecies interactions}

Let us now consider the unequal intraspecies interactions, with which
there could be an imbalanced in the species population, given by $x=\phi_{2c}/\phi_{1c}=\sqrt{n_{2}/n_{1}}$.
Taking the small chemical potential limit as in the case of equal
intraspecies interactions, i.e., $\mu_{1}=\mu_{2}=0$ in $\varOmega_{\textrm{LHY}}$,
we find that,
\begin{eqnarray}
C_{1} & = & x\Delta,\\
C_{2} & = & x^{-1}\Delta,\\
B_{1\mathbf{k}} & = & \frac{\hbar^{2}\mathbf{k}^{2}}{2m}+2x\Delta,\\
B_{2\mathbf{k}} & = & \frac{\hbar^{2}\mathbf{k}^{2}}{2m}+2x^{-1}\Delta.
\end{eqnarray}
By introducing the variable $t=[\hbar^{2}k^{2}/(2m)]/(2\Delta)$,
we can write $\varOmega_{\textrm{LHY}}$ into the form,
\begin{equation}
\varOmega_{\textrm{LHY}}=\frac{32\sqrt{2}m^{3/2}}{15\pi^{2}\hbar^{3}}\Delta^{5/2}f\left(x\right),
\end{equation}
where the function is defined by,
\begin{equation}
f\left(x\right)=\frac{15\sqrt{2}}{32}\int_{0}^{\infty}dt\sqrt{t}\left[\tilde{E}_{+}\left(t\right)+\tilde{E}_{-}\left(t\right)-\left(2t+x+x^{-1}\right)+\frac{\left(x+x^{-1}\right)^{2}}{8t}\right],
\end{equation}
with 
\begin{equation}
\tilde{E}_{\pm}^{2}\left(t\right)=\frac{\left(\tilde{B}_{1t}^{2}-\tilde{C}_{1}^{2}\right)+\left(\tilde{B}_{2t}^{2}-\tilde{C}_{2}^{2}\right)^{2}}{2}-\frac{1}{4}\pm\frac{1}{2}\sqrt{\left[\left(\tilde{B}_{1t}^{2}-\tilde{C}_{1}^{2}\right)-\left(\tilde{B}_{2t}^{2}-\tilde{C}_{2}^{2}\right)^{2}\right]^{2}+\left(\tilde{C}_{1}+\tilde{C}_{2}\right)^{2}-\left(\tilde{B}_{1t}-\tilde{B}_{2t}\right)^{2}}.
\end{equation}
Here, we have introduced the notations: $\tilde{C}_{1}=x/2$, $\tilde{C}_{2}=x^{-1}/2$,
$\tilde{B}_{1t}=t+x$ and $\tilde{B}_{2t}=t+x^{-1}$.

\begin{figure}[t]
\begin{centering}
\includegraphics[width=0.5\textwidth]{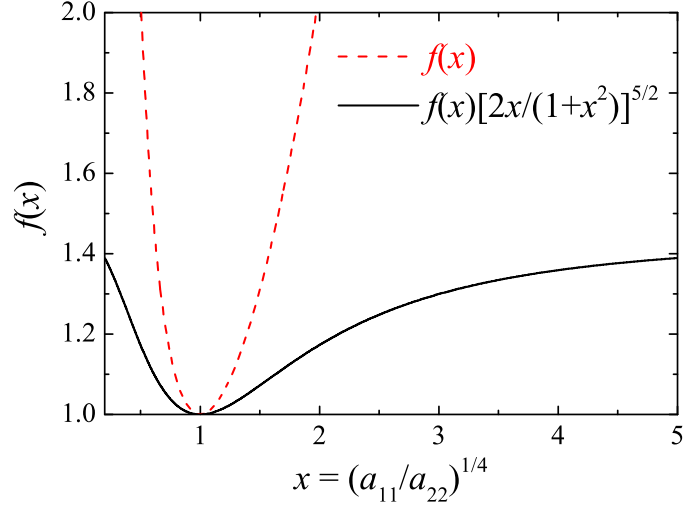}
\par\end{centering}
\caption{\label{figSM5_fx} The function $f(x)$ and the enhancement factor
$f(x)[2x/(1+x^{2})]^{5/2}$ as a function of $x=(a_{11}/a_{22})^{1/4}$.
We note that, both functions are symmetric with respect to the point
$x=1$, i.e., $f(x)=f(x^{-1}).$}
\end{figure}

By adding $\varOmega_{0}$, we obtain at the unequal intraspecies
interactions, 
\begin{equation}
\varOmega\simeq-\frac{m}{8\pi\hbar^{2}}\left[\frac{\mu_{1}x\Delta}{a_{11}}+\frac{\mu_{2}x^{-1}\Delta}{a_{22}}+\frac{x^{2}\Delta^{2}}{a_{11}}+\frac{x^{-2}\Delta^{2}}{a_{22}}+\frac{2\Delta^{2}}{a_{12}}\right]+\frac{32\sqrt{2}m^{3/2}}{15\pi^{2}\hbar^{3}}\Delta^{5/2}f\left(x\right).
\end{equation}
Taking the saddle point $\Delta=\Delta_{0}$ and the derivative of
$\varOmega(\Delta_{0})$with respect to $\mu_{1}$ and $\mu_{2}$,
we find that,
\begin{eqnarray}
n_{1} & \simeq & \frac{xm\Delta_{0}}{4\pi\hbar^{2}a_{11}},\\
n_{2} & \simeq & \frac{x^{-1}m\Delta_{0}}{4\pi\hbar^{2}a_{22}}.
\end{eqnarray}
By dividing these two expressions with each other, we find that 
\begin{equation}
x^{2}=\frac{n_{2}}{n_{1}}=\sqrt{\frac{a_{11}}{a_{22}}}.
\end{equation}
This identity has also obtained in Petrov's theory, although a quite
different derivation (i.e., starting from the mean-field energy, which
is different from ours) is demonstrated. The coincidence is interesting.
We can replace the pairing parameter $\Delta_{0}$ by the density,
i.e., 
\begin{equation}
\Delta_{0}=\frac{4\pi\hbar^{2}\sqrt{a_{11}a_{22}}}{m}\left(n_{1}n_{2}\right)^{/12}=\frac{4\pi\hbar^{2}a}{m}\left(n_{1}n_{2}\right)^{/12}.
\end{equation}
By calculating the total energy $E=\varOmega+\mu_{1}n_{1}+\mu_{2}n_{2}$,
we obtain,
\begin{equation}
\frac{E}{N}=-\frac{\pi\hbar^{2}}{m}\left(a+\frac{a^{2}}{a_{12}}\right)\left[\frac{2x}{1+x^{2}}\right]^{2}n+\frac{256\sqrt{\pi}}{15}\frac{\hbar^{2}a^{5/2}}{m}\left[\frac{2x}{1+x^{2}}\right]^{5/2}f\left(x\right)n^{3/2}.
\end{equation}
This energy is to be compared with Petrov's prediction \cite{Petrov2015},
\begin{equation}
\frac{E_{\textrm{Petrov}}}{N}=\frac{\pi\hbar^{2}}{m}\left(a+a_{12}\right)\left[\frac{2x}{1+x^{2}}\right]^{2}n+\frac{256\sqrt{\pi}}{15}\frac{\hbar^{2}a^{5/2}}{m}n^{3/2}.
\end{equation}
We emphasize that in our pairing theory, the LHY energy term is enhanced
by a factor of 
\begin{equation}
\eta\left(x\right)=\left[\frac{2x}{1+x^{2}}\right]^{5/2}f\left(x\right),
\end{equation}
which could be very significant for a \emph{large} imbalance in intraspecies
interactions. For example, if $x=2$ (or $x=0.5$) at $a_{11}=16a_{22}$
(or $a_{11}=a_{22}/16$), the enhancement factor can be around $\eta\simeq1.2$
and hence decrease the equilibrium density by a factor of $\eta^{2}\simeq1.4$.
In the current experiments of a $^{39}$K Bose-Bose mixture, the ratio
of $a_{11}/a_{22}$ is about $2$ and then $x\simeq1.2$. Therefore,
the enhancement in the LHY energy is just a few percent.

\end{widetext}
\end{document}